\documentclass[aps,twocolumn,reprint,superscriptaddress,floatfix,longbibliography]{revtex4-2}%
\usepackage{float}
\usepackage{graphicx}
\usepackage{amsmath,amssymb,bbm,mathrsfs,bm,braket,color,graphicx,comment,amsfonts,dsfont}
\usepackage[colorlinks,citecolor=blue,urlcolor=blue]{hyperref}
\usepackage[mathscr]{euscript}
\usepackage[normalem]{ulem}
\usepackage{comment}

\DeclareMathAlphabet\mathbfcal{OMS}{cmsy}{b}{n}

\usepackage{extarrows} 
\usepackage{mathrsfs}

\begin{document}

\title{Electrically Controllable Landau Levels in Two-Dimensional Electron Gases \\ under Nonuniform Magnetic Fields }

\author{You-Ting Huang}
\affiliation{Research Center for Applied Sciences, Academia Sinica, Taipei 11529, Taiwan}

\author{Chao-Cheng Kaun} 
\email{kauncc@gate.sinica.edu.tw}
\affiliation{Research Center for Applied Sciences, Academia Sinica, Taipei 11529, Taiwan}

\author{Ching-Hao Chang}
\email{cutygo@phys.ncku.edu.tw}
\affiliation{Department of Physics, National Cheng Kung University, Taiwan}
\affiliation{Center for Quantum Frontiers of Research and Technology (QFort), National Cheng Kung University, Tainan 70101, Taiwan}

\begin{abstract}

Flat bands underlie a diverse range of quantum phenomena, from strongly correlated phases to superconductivity. We theoretically establish that a two-dimensional electron gas under a linear magnetic-field gradient and a transverse electric field exhibits electrically tunable flat bands. When the electric field magnitude is tuned to a value within a discrete sequence, these bands become strictly dispersionless. By providing exact classical and quantum solutions, we demonstrate that these states are high-order Landau levels associated with drift-compensated cyclotron orbits of carriers arising from the synergy between the magnetic-field gradient and the electric field. These electrically controllable Landau levels exhibit quantized Hall conductance and a strongly enhanced density of states. Our results provide a new route for flat-band creation, magnetoelectric band engineering, and quantized Hall currents controlled via source-drain voltage.
\end{abstract}

\maketitle
\textcolor{blue}{\textit{Introduction}} --- 
In modern condensed matter physics, several research directions are at the forefront of current interest, including superconductivity\cite{superconductivity1,superconductivity2,superconductivity3}, the quantum Hall effect\cite{Hall1,Hall2,Hall3,BoltzHall}, and quantum information science\cite{information1,information2}. Many of these high–application-value phenomena are closely connected to an exotic class of electronic dispersion known as flat bands\cite{flatband1}.
Flat bands strongly enhance the electronic density of states, thereby amplifying interaction effects: even weak electron–electron interactions can induce symmetry breaking, correlated insulating phases, or superconductivity\cite{flatband2}. Moreover, owing to their quenched kinetic energy, flat bands are exceptionally sensitive to external perturbations such as electromagnetic fields or strain, providing a powerful route toward tunable electronic properties\cite{flatband3}.

The most fundamental and historically earliest realization of flat bands arises from the case of uniform magnetic field, which quantizes the electronic spectrum into Landau levels with harmonic-oscillator–like energies 
$E_n=(n+1/2)\hbar \omega_c$
, where
$\omega_c=qB/m^\star$
is the cyclotron frequency and $m^\star$ denotes the effective mass of carriers\cite{Kittel,plasma}. 
Beyond this canonical mechanism, a variety of innovative approaches have been proposed to engineer flat bands, including destructive interference through special lattice geometries \cite{flatband4,flatband5}, strain and twist engineering of two-dimensional materials such as magic-angle twisted bilayer graphene \cite{flatband6}, and periodic orbital magnetic fields that generate emergent flat-band lattices at magic field values \cite{Magicfield}. The physics of flat bands in tight-binding networks — from compact localized states and singular band-touching to nonlinear caging and many-body nonergodic models — has been systematically developed \cite{Danieli2021,Danieli2024}, and these ideas have been actively explored in electronic, photonic, and quantum platforms \cite{Poblete2021,flatband7}.



While these approaches have achieved impressive progress, their dependence on specific lattice geometries, twist angles, or strain profiles limits material generality and tunability. In this study, we propose a conceptually distinct route to flat bands through electric-field control in two-dimensional electron gases (2DEGs) under nonuniform magnetic fields. Begin with analyzing the electronic structure of 2DEGs subjected to a linearly varying magnetic-field dipole. The resulting band structure and wave functions exhibit pronounced time-reversal asymmetry and a characteristic unidirectional group-velocity behavior on both sides of the dipole\cite{PhysRevLett1992}. 
Upon introducing a uniform transverse electric field, a group-velocity imbalance emerges and lifts the degeneracy between left- and right-localized modes. Remarkably, at a sequence of discrete electric field strengths, different energy levels become completely flat, realizing electrically controllable, Landau-level–like quantization in a nonuniform magnetic field. Furthermore, at a precisely tuned electric-field strength, the Hamiltonian becomes exactly solvable, and the resulting closed-form solution coincides with the classical condition under which the
$\nabla B$
drift and the
$\rm\textbf{E}\times\rm\textbf{B}$
drift cancel. 
These results establish a new mechanism for magneto-electric band engineering and guided electronic transport in low-dimensional systems under nonuniform magnetic fields.

\begin{figure}[h!]
    \centering
    \includegraphics[width=.95\linewidth]{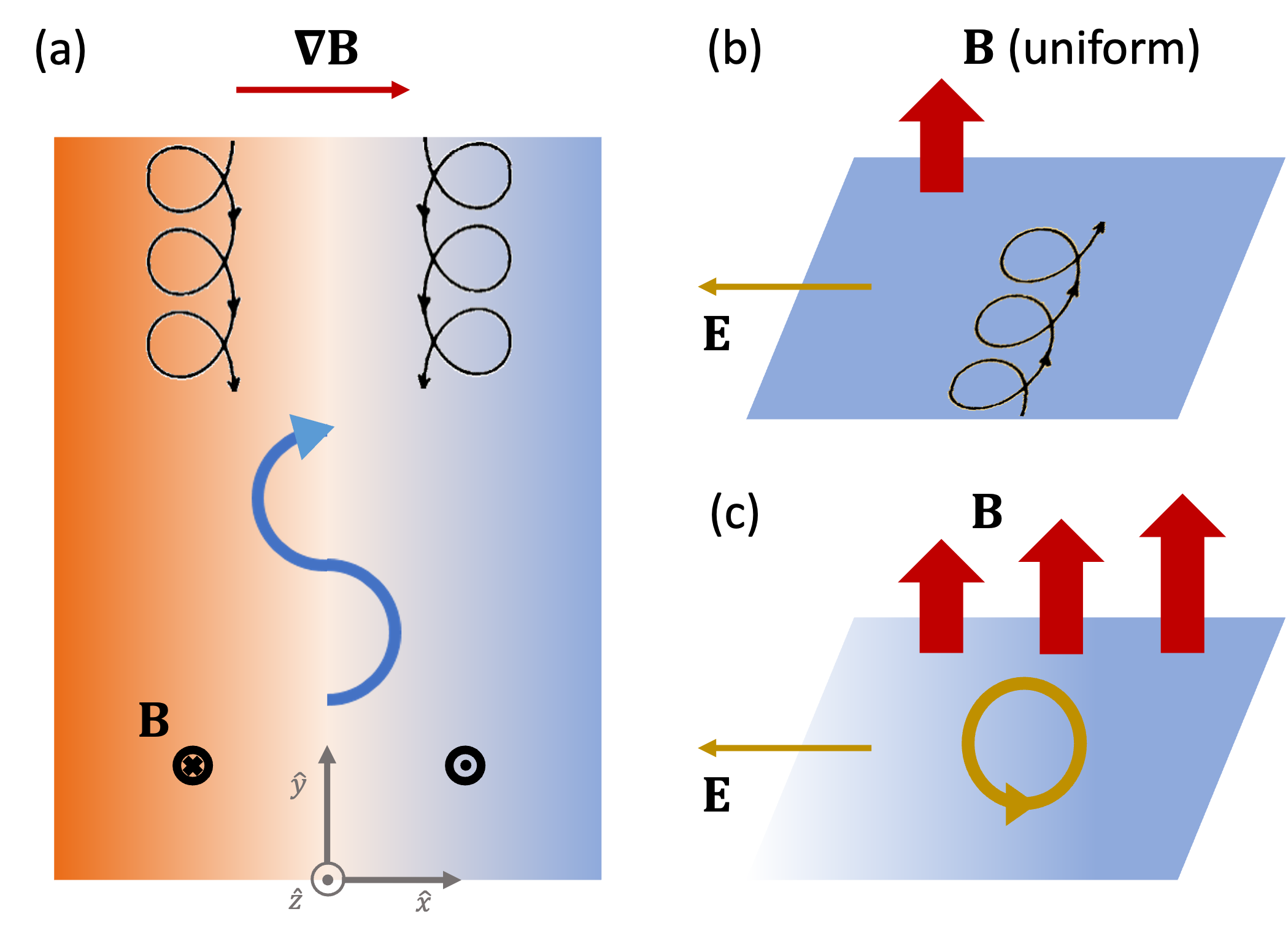}
    \caption{Classical pictures of electron motions in different electromagnetic field configurations: (a) a linearly varying magnetic-field dipole spans from negative (into the paper) to positive (out of the paper); (b) a uniform magnetic field with a uniform electric field perpendicular to it; (c) a linear gradient magnetic field with a uniform electric field perpendicular to it (and parallel to its gradient direction.)}
    \label{fig1}
\end{figure}

\textcolor{blue}{\textit{The classical model}} --- 
To construct a clear physical picture, let’s begin with the classical model. Consider a two-dimensional electron gas (2DEG) on the x–y plane. The magnetic field $B(x)\hat{z}$ varies linearly along $\hat{x}$, and a uniform transverse electric field is applied along $\hat{x}$ (parallel to the magnetic field gradient and perpendicular to $\textbf{B}$).
The dynamics of a charged carrier in electromagnetic fields are governed by the Lorentz equation
\[\rm\textbf{F}=q(\rm\textbf{E}+\textbf{v}\times\textbf{B})\] 
In a uniform magnetic field, charged carriers execute circular cyclotron motion, with the curvature radius defines the Lamor radius
$r_L= v/\omega_c$
, where $v$ is the velocity of charged carriers.
For a linearly varying magnetic-field dipole
\begin{align}
    \textbf{B}=B(x)\hat{z}=\frac{B_0x}{L}\hat{z}
    \label{eq1}
\end{align}
, with $L$ a length scale parameter such that $|\textbf{B}|=B_0$ as $x=\pm L$, charged carriers' curvature radius depends on position through the local field magnitude.  As a result, the guiding center of the cyclotron orbit acquires a drift velocity known as the gradient B ($\nabla B$) drift,\cite{plasma}
\begin{align}
    \textbf{v}_{\nabla B}=\operatorname{sign}(q) \frac{1}{2} r_Lv\frac{\textbf{B}\times\nabla B}{B^2}
    \label{eq2}
\end{align}

Around the interface where the magnetic field changes sign, the orbital chirality of cyclotron motion reverses across the zero-field line. This leads to the formation of snake orbits\cite{ChingHao,ChangHao2}, which propagate along the interface perpendicular to  $\nabla B$.


Another important classical scenario is a uniform magnetic field combined with a uniform electric field perpendicular to it, as shown in Fig.~\ref{fig1}(b). In this case, electrons undergo the $\textbf{E}\times \textbf{B}$ drift, with the drift velocity\cite{plasma}
\begin{align}
    \textbf{v}_{\textbf{E}\times \textbf{B}}=\frac{\textbf{E}\times \textbf{B}}{B^2}
    \label{eq3} 
\end{align}
, which plays a key role in the classical Hall effect.


Next, we consider a more general field configuration consisting of a linear gradient magnetic field, together with a uniform electric field that is perpendicular to the magnetic field and parallel to its gradient, as illustrated in Fig.~\ref{fig1}(c). In this case, the total drift velocity is given by
\[\textbf{v}_{d}=\textbf{v}_{\nabla B}+\textbf{v}_{\textbf{E}\times\textbf{B}}\]
Consider $q=-e$, from Eq.(\ref{eq1}), Eq.(\ref{eq2}), and Eq.(\ref{eq3}), we deduce the critical electric field
\begin{align}
    \textbf{E}_c=-\frac{1}{2}\frac{m^\star v^2}{eL}\hat{x}
    \label{eq4}
\end{align}
at which the $\nabla B$ drift and the $\textbf{E}\times\textbf{B}$ drift cancel each other. Under this condition, the guiding-center drift vanishes, restoring zero net transport and resulting in a drift-compensated cyclotron orbit.

In this study, we find that, within the quantum-mechanical framework, the Hamiltonian becomes analytically solvable under this same field configuration. Remarkably, the resulting closed-form quantum solution corresponds perfectly to the classical drift-cancellation condition, establishing a direct connection between classical electromagnetic dynamics and the emergence of flat, Landau-level-like energy bands.

\textcolor{blue}{\textit{Hamiltonian and numerical results}}---
In quantum mechanics, the behavior of electorns in electromagnetic fields is governed by the Hamiltonian
\[\textbf{H}=\frac{(\textbf{p}+e\textbf{A)}^2}{2m^\star}+E_e\]
To establish the field configuration discussed above, we chose the gauge vector as
$\textbf{A}=(B_0x^2/2L)\hat{y}$
such that its curl yields the linearly varying magnetic-field dipole consistent with Eq.(\ref{eq1}). Since the Hamiltonian is translationally invariant along the $y$ direction, the canonical momentum $p_y$ can be replace by its eigenvalue $hk_y$. The electric field is incorporated through the electric potential energy
$E_e=(eV_e/L)x$
which corresponds to a uniform electric field that satisfies $E_e=\pm eV_e$ at $x=\pm L$.
Under these conditions, the Hamiltonian describing our system, a 2DEG subjected to a linearly varying magnetic-field dipole together with a uniform transverse electric field, can be written as
\begin{align}
    \textbf{H}=\frac{\textbf{p}_x^2}{2m^\star}+V(x)
    \label{eq5}
\end{align}
with the quartic effective potential
\[V(x)=\frac{1}{2m^\star}(\hbar k_y+\frac{m^\star\omega_c}{2L}x^2)^2+\frac{eV_e}{L}x\]
In general, it does not admit a closed-form analytical solution\cite{anharmonic}. We therefore proceed by employing a numerical approach based on the operator formalism of the quantum harmonic oscillator.


We expand the Hamiltonian in the basis of Hermite functions. Specifically, the power terms of position operator $x$, $x^2$, and $x^4$ can be expressed in terms of the harmonic oscillator ladder operators $\hat{a}$ and $\hat{a}^\dagger$. Using the relation between the ladder operators and the Hermite basis\cite{griffiths_introduction_2018}, the matrix representation of the Hamiltonian Eq.(\ref{eq5}) can be deduced. The resulting matrix can then be efficiently diagonalized numerically using Mathematica, yielding the energy spectrum and corresponding eigenstates. The effective mass is chosen as
$m^\star=0.067\rm m_e$
to simulate the GaAs/AlGaAs heterostructure\cite{2DEG1,2DEG2}; the magnetic field strength parameter is
$B_0=1.65\rm T$
, and the length scale is
$L=16\pi^2\rm nm$.
The results are evaluated in the ballistic regime, assuming the mean free path of carriers is longer than the system length, so that the transmission is determined entirely by the band structure without scattering.\cite{ballistic}

\begin{figure}[h!]
    \centering
    \includegraphics[width=1.\linewidth]{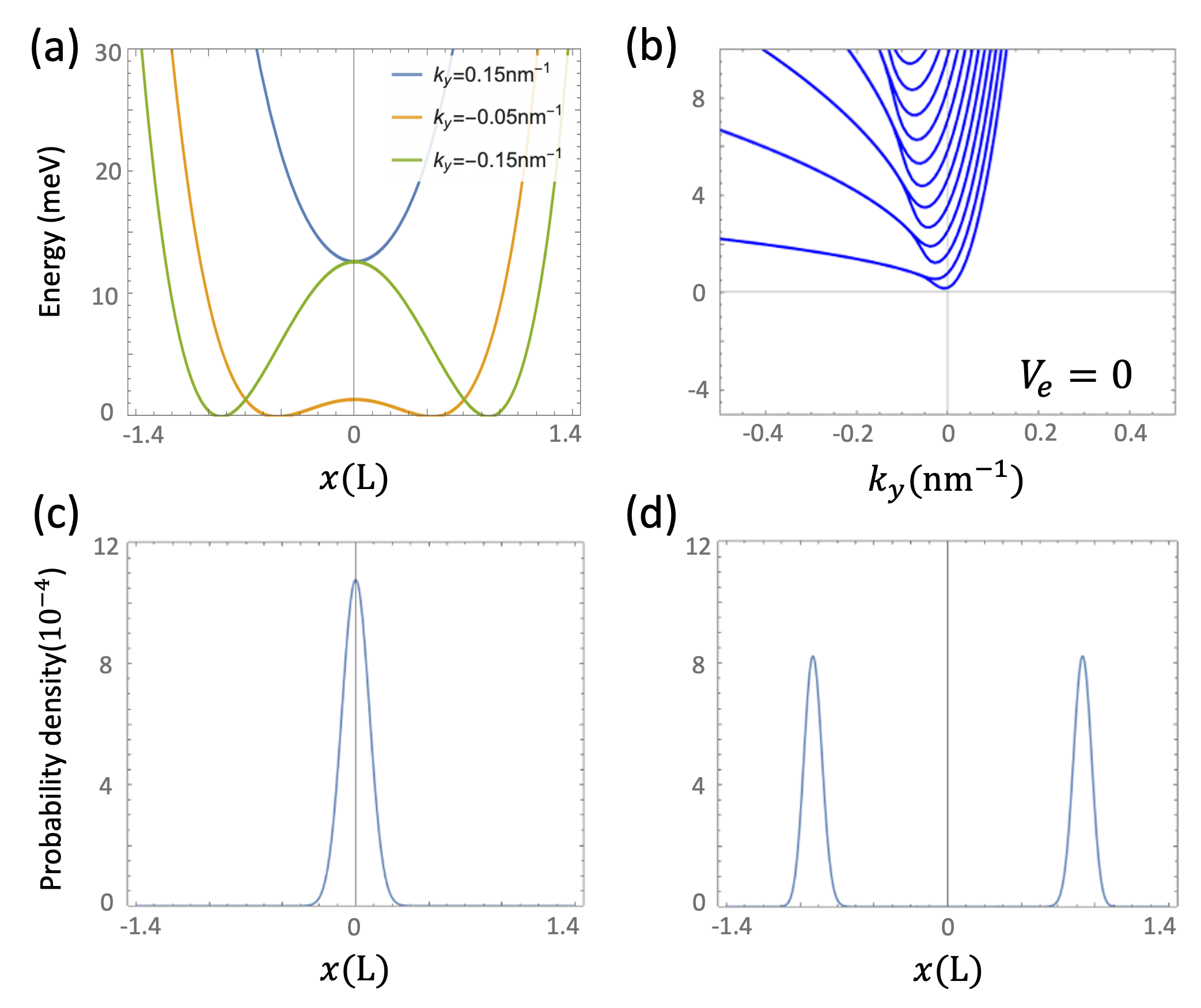}
    \caption{Colored line. Numerical results in the absence of electric field. (a) the effective potential $V(x)$ for $k_y=0.15\rm nm^{-1}$(Blue), $k_y=-0.05\rm nm^{-1}$(Orange), and $k_y=-0.15\rm nm^{-1}$(Green); (b) the energy spectrum; (c) and (d) the probability density for $k_y=0.15\rm nm^{-1}$ and $k_y=-0.15\rm nm^{-1}$. $L$ is set to $16\pi^2\rm nm$ in the computation.}
    \label{fig2}
\end{figure}


Figure~\ref{fig2} shows the energy spectrum in the absence of an electric field ($V_e=0$), together with the corresponding effective potentials and probability densities for representative values of $k_y$. For $k_y>0$, the effective potential exhibits a single-well, free-electron–like profile, leading to wave functions localized near the interface where the magnetic field changes sign. These states possess positive group velocity and correspond to snake orbits in the classical picture.
In contrast, for $k_y<0$, the effective potential develops a double-well structure, resulting in states that are localized on both sides of the magnetic-field dipole. These double degenerate modes exhibit negative group velocity and correspond to the classical $\nabla B$ drift. Each pair of degenerate bands splits into two branches as $k_y$ varies from negative to positive values due to the $k_y$-dependence of the effective potential (see Fig. \ref{fig2} (a)). The asymmetry of the energy spectrum reflects the pronounced breaking of time-reversal symmetry.

To further examine the nature of the negative-$k_y$ states, we expand the effective potential $V(x)$ around its minimum points
\[x_0=\pm\sqrt{\frac{-2\hbar k_yL}{m^\star\omega_c}}\]
The effective potential to quadratic order can be approximated as
\begin{align}
    V(x)\approx\frac{1}{2}m^\star\omega_c^2\frac{x_0^2}{L^2}(x-x_0)^2
    \label{eq6}
\end{align}
which is almost identical to the harmonic potential of Landau levels. The only difference is that the cyclotron frequency $\omega_c$ is replaced by an effective local cyclotron frequency $\omega_c(x_0/L)$, determined by the position of the potential minimum. Within this low-energy approximation, the double-well potential can therefore be regarded as a combination of two harmonic potentials centered at 
$\pm x_0$
, separated by an energy barrier at the center. The corresponding eigenvalue can be approximated as
\begin{align}
    E_n\approx\hbar\omega_c\frac{|x_0|}{L}(n+\frac{1}{2})
    \label{eq7}
\end{align}
which explains the observed dispersion relation $E_n\propto\sqrt{-k_y}$ and the double degeneracy for $k_y<0$. 

From Eq.~(\ref{eq7})
and the definition of group velocity $v_g=(1/\hbar)(\partial E/\partial k)$, we obtain the group velocity of the degenerate states
\begin{align}
    \textbf{v}_g=\frac{-\hbar\omega_c}{eB_0|x_0|}(n+\frac{1}{2})\hat{y}
    \label{eq8}
\end{align}
On the other hand, evaluating the classical $\nabla B$ drift velocity from Eq. (\ref{eq2}) at the local field $\textbf{B}(x_0)=(B_0x_0/L)\hat{z}$ gives
\begin{align}
    \textbf{v}_{\nabla B}=\frac{-\frac{1}{2}m^\star v^2}{eB_0|x_0|}\hat{y}
    \label{eq9}
\end{align}  
A direct quantum-classical correspondence is thus established: the group velocity of the degenerate quantum states (Eq.~(\ref{eq8})) coincides with the classical $\nabla B$ drift velocity (Eq.~(\ref{eq9})), with the Landau level eigenenergy $\hbar\omega_c(n+1/2)$ replaced by the classical kinetic energy $(1/2)m^\star v^2$.

\begin{figure}[h!]
    \centering
    \includegraphics[width=1.\linewidth]{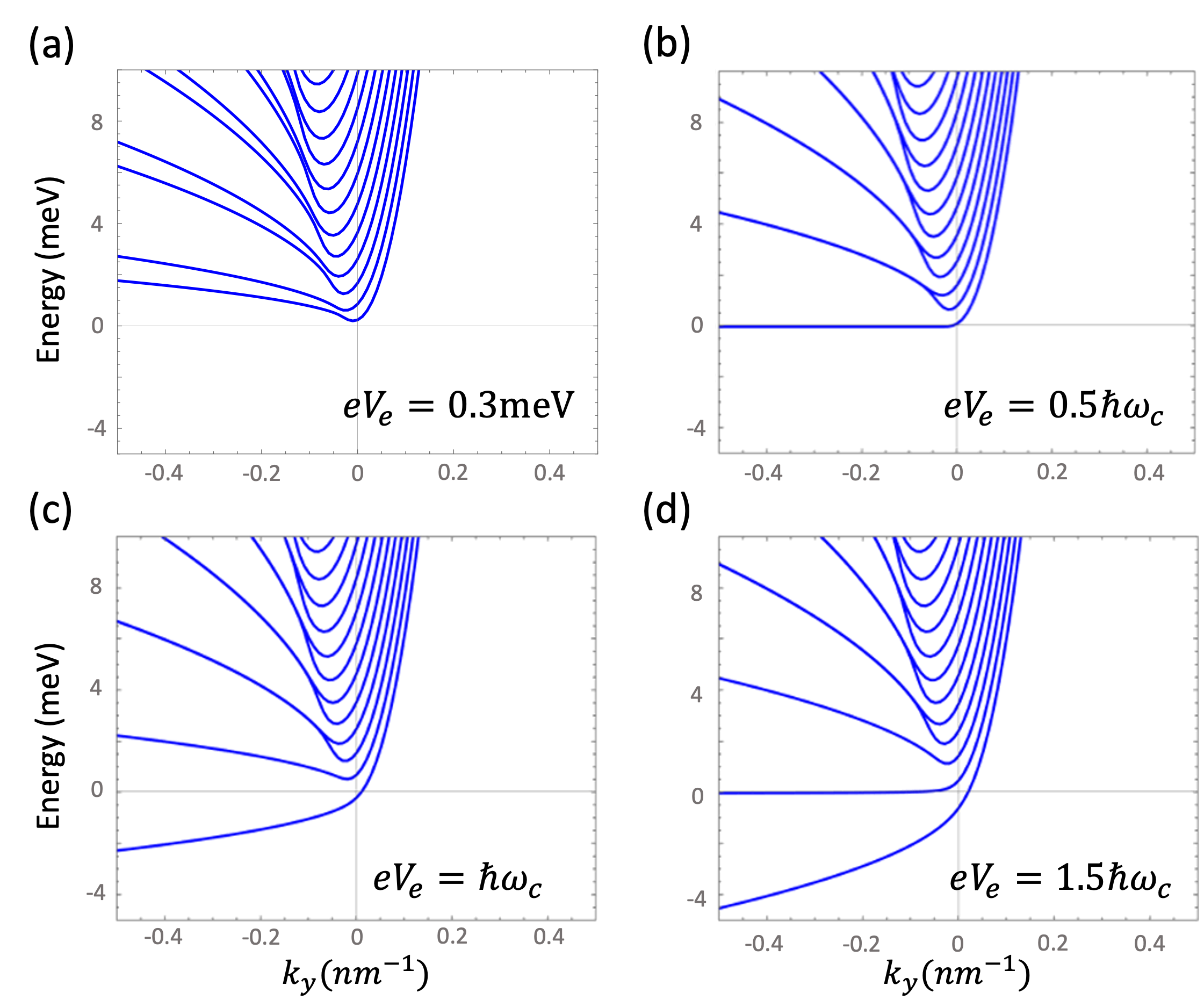}
    \caption{Energy spectra for 2DEG subjected to the linearly varying magnetic-field dipole, together with uniform electric fields of different strengths that are perpendicular to the magnetic field and parallel to its gradient. The corresponding electric potential energies $eV_e$ are: (a) $0.3\rm meV$, (b) $0.5\hbar\omega_c$, (c) $\hbar\omega_c$, and (d) $1.5\hbar\omega_c$ }
    \label{fig3}
\end{figure}

Next, we consider explicitly the effects of electric fields. Figure~\ref{fig3} presents the energy spectra for several representative electric-field strengths. The electric field enters the Hamiltonian as a linear electric potential term, which tilts the double-well potential and give opposite energy shifts $\Delta E_e=\pm eV_e(x_0/L)$ at two minima $\pm x_0$, respectively. Therefore, as shown in Fig.~\ref{fig3}(a), the degeneracy of the double-well states at $k_y<0$ is lifted when a weak electric field is applied, while the snake states at $k_y>0$ remain almost unchanged. Interestingly, in Fig.~\ref{fig3}(b), when $eV_e$ is tuned to the ground state energy of Landau levels, $(1/2)\hbar\omega_c$, one of the previously degenerate band becomes completely flat (the calculated eigenvalue variations is seven orders of magnitude smaller than the band gap), while the remaining bands recover degeneracy. This behavior can be understood as follows: the energy of one localized ground state is canceled by the electric potential energy $E_e=0.5\hbar\omega_c(x_0/L)$ whereas the other is raised to $\hbar\omega_c$, coinciding with the first excited state lowered by the same $E_e$.

Figures~\ref{fig3}(c) and (d) demonstrate that the phenomenon occurs at stronger electric field strengths: when the electric field is tuned to $eV_e = n\hbar\omega_c$ ($n$ an integer), new degeneracies emerge while $2n$ bands become nondegenerate\cite{note1}; when it is tuned to $eV_e = (n+1/2)\hbar\omega_c$, flat bands appear. These results establish an electrically switchable mechanism for flat-band formation, offering enhanced tunability and broad material generality across low-dimensional electron systems.

It is instructive to compare this electrical controllable Landau level with the standard ones in a uniform magnetic field. There, the spectrum $E_n = (n + 1/2)\hbar \omega_c$ is independent of $k_y$. Cyclotron orbits are closed and the guiding-center drift vanishes identically because no mechanism exists to displace the guiding center, and flat bands therefore holds automatically for every $n$. In our nonuniform-field system, flatness is not automatic: it is achieved only when the applied electric field is tuned so that the $\textbf{E}\times \textbf{B}$ drift exactly cancels the $\nabla B$ drift, restoring the zero-drift condition. The mechanism in both cases is analogous — flat bands correspond to vanishing guiding-center drift — but our system requires the interplay of the electromagnetic fields: cancellation occurs only at the discrete sequence of electric field strengths, occurring for only single state with zero energy, providing new avenues for external field control.

\textcolor{blue}{\textit{The analytical solution}}---
Although solving Eq.(\ref{eq5}) generally requires numerical methods, it becomes analytically solvable in a special case. For the flat band shown in Fig.~\ref{fig3}(b), we assume the eigenenergy to be exact zero, then the underlying Schrödinger equation reads
\begin{align}
    \frac{\textbf{p}_x^2}{2m^\star}\Psi(x)+(\frac{1}{2m^\star}(\hbar k_y+\frac{m^\star\omega_c}{2L}x^2)^2+\frac{\hbar\omega_c}{2L}x)\Psi(x)=0
    \notag\\
    \label{eq10}
\end{align}
for $k_y<0$, where the wave function admit a closed-form analytical solution
\begin{align}
    \Psi(x)=A{\rm exp}[k_yx+\frac{m^\star\omega_c}{6\hbar L}x^3]
    \label{eq11}
\end{align}
with $A$ a normalization constant. 
This wave function is normalizable only on the interval $(-\infty,0]$, indicating that the physically valid bound state is localized on the lower side of the tilted double-well potential. 
The critical electric field leading to the solvable condition is
\begin{align}
    \textbf{E}_c=-\frac{1}{2}\frac{\hbar\omega_c}{eL}\hat{x}
    \label{eq12}
\end{align}
which coincides with Eq.(\ref{eq4}), the classical condition under which the $\nabla B$ drift and the $\textbf{E}\times\textbf{B}$ drift cancel, with the classical kinetic energy $(1/2)m^\star v^2$ replaced by $(1/2)\hbar\omega_c$, the ground state energy of Landau levels. Upon reversing the electric field, the wave function becomes 
\[\Psi_{-\textbf{E}}(x)=A{\rm exp}[-k_yx-\frac{m^\star\omega_c}{6\hbar L}x^3]\]
implying that the double-well potential is tilted in the opposite direction and that the bound state is localized near the other potential minimum.


Based on the numerical results, flat bands appear not only under the condition of Eq.(\ref{eq12}), but at a sequence of discrete electric fields $\textbf{E}=(n+1/2)\hbar\omega_c/(eL)\hat{x}$. To construct the solutions for flat bands with different $n$ (called excited state), we follow an approach analogous to that used for excited states of the quantum harmonic oscillator and adopt the ansatz\cite{griffiths_introduction_2018}
\[\Psi_n(x)=P_n(x)\Psi(x)\]
Substituting it into the Schrödinger equation yields the following differential equation for $P_n(x)$
\begin{align}
    P_n''(x)+2f(x)P_n'(x)-2nf'(x)P_n(x)=0
    \label{eq13}
\end{align}
where 
\[f(x)=k_y+\frac{m^\star\omega_c}{2\hbar L}x^2\]
If $f(x)$ were replaced by $-x$, Eq.(\ref{eq11}) would reduce exactly to the Hermite differential equation
\cite{math_for_physicist}, whose solutions are Hermite polynomials. However, for the present case $f(x)\sim x^2$, Eq.(\ref{eq11}) admits no closed-form solution. Nevertheless, Eq.(\ref{eq11}) provides a numerically advantageous formulation by separating the known envelope from the polynomial structure, enabling a stable construction of the excited states of the electrically switchable flat bands.

\begin{figure}[h!]
    \centering
    \includegraphics[width=1.\linewidth]{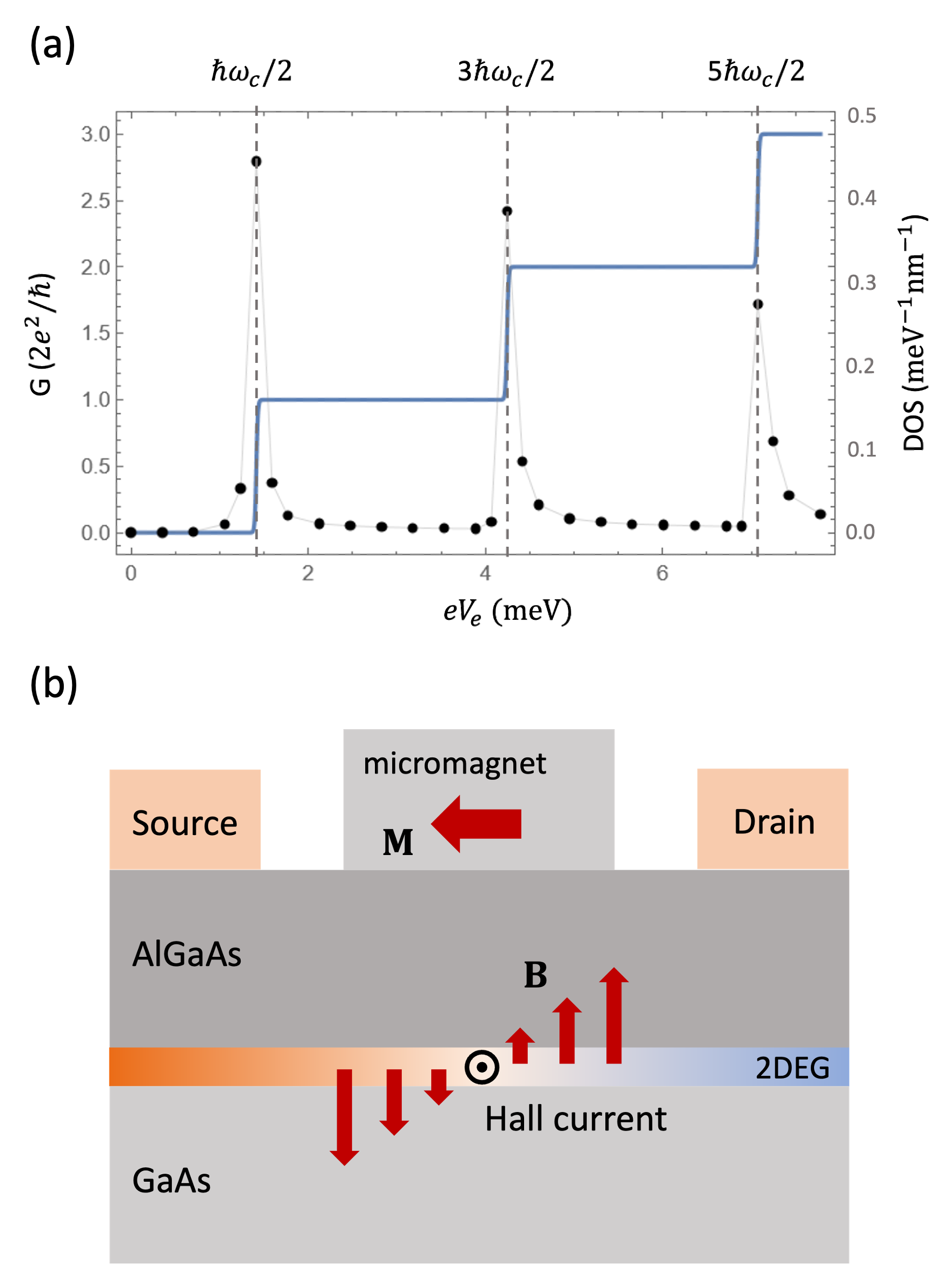}
    \caption{(a) Hall conductance along $\hat{y}$ (blue line, left axis) and density of states at the Fermi level $E_F \approx 0$ (black dots, right axis) as functions of longitudinal electric field strength, corresponding to the source-drain voltage in the nano-electronic device in panel.
    (b) Schematic diagram of the experiment setup. The periodic arranged magnets generate periodic effective magnetic field, which can be approximated as the linearly varying magnetic-field dipole.}
    \label{fig4}
\end{figure}

\textcolor{blue}{\textit{Quantized Hall conductance induced by source-drain voltage}}---
Flat bands exhibit characteristic transport signatures that are expected to give rise to distinct physical phenomena. Figure~\ref{fig4} shows the conductance along $\hat{y}$ and the density of states at the Fermi energy $E_F=0$ as functions of the transverse electric field strength $eV_e$.
The conductance is evaluated using the Landauer formula\cite{Landauer} at temperature $T=0.1\rm{K}$. As $eV_e$ reaches $(n+1/2)\hbar\omega_c$, the conductance increases in a step-like manner. This behavior can be understood from the band structure shown in Fig.~\ref{fig3}: when $eV_e$ exceeds $(n+1/2)\hbar\omega_c$, $n+1$ bands that originally possess negative slopes become positively sloped. This corresponds to the classical condition under which the $\mathbf{E}\times\mathbf{B}$ drift surpasses the $\nabla B$ drift, causing electrons to propagate along the same direction as the snake states and thereby enhancing the net conductance.

To evaluate the density of states, we introduce a Gaussian broadening function\cite{math_for_physicist}
\[{\rm exp}[-\frac{(E_F-E_n)^2}{\eta^2}]\]
to regularize the divergence associated with flat bands. The broadening parameter $\eta$ accounts for the effects of finite temperature and disorder\cite{Kittel} and is chosen to be $0.1\rm{meV}$. Pronounced peaks in the density of states emerge as $eV_e$ reaches $(n+1/2)\hbar\omega_c$,  reflecting the strong density of states enhancement associated with the electrically switchable flat band.

Experimentally, the linearly varying magnetic-field dipole in 2DEGs can be realized by a ferromagnetic micromagnet strip with in-plane magnetization perpendicular to the channel direction, fabricated on top of the GaAs/AlGaAs heterostructure \cite{Roukes1987,Dweiss1997,Dweiss1995}. The micromagnet generates a prominent magnetic-field dipole in the 2DEG, with snake states experimentally observed at the zero-magnetic-field crossing \cite{expB}. A uniform transverse electric field can then be applied via source-drain biasing or electrostatic gating \cite{expE}. This combination of a locally linear magnetic gradient and a uniform electric field offers an experimentally accessible platform for designing a Hall transistor controlled by the source-drain voltage (see Fig.~\ref{fig4}).

Another realization of our field configuration is by applying a uniform magnetic field to a curved nanoribbon together with an electric potential difference between the two edges. The effective magnetic field is the component perpendicular to the ribbon's surface, which varies sinusoidally along the ribbon width and can be approximated as linear to leading order across the sign-change region \cite{PRB_205309,BoltzHall,ChingHao}. The smooth magnetic-field dipole profile in such systems can be approximated as linear only near its zero crossing. As a result, the $\nabla B$ drift can only be compensated locally by the uniform $\mathbf{E}\times\mathbf{B}$ drift, causing the magnetic bands to acquire a finite dispersion rather than remain perfectly flat. Nevertheless, the quantized Hall effect is expected to remain experimentally observable because the electric-field-induced evolution of the magnetic band structure shown in Fig.~\ref{fig3} is preserved approximately, although the Hall plateaus may become less sharp than in the ideal linear-gradient case.

\textcolor{blue}{\textit{Summary}} --- 
In this study, we calculate the energy spectra and wave functions of a 2DEG subjected to a linearly varying magnetic-field dipole together with a uniform transverse electric field, using the operator formalism of the quantum harmonic oscillator. In the absence of an electric field, the energy spectra exhibit pronounced time-reversal asymmetry due to the dependence of the quartic effective potential on the momentum  $k_y$. When electric fields are included, flat bands emerge at a sequence of discrete electric field strengths. We further obtain an analytical solution for the ground-state flat band, which corresponds to the classical drift cancellation condition, while excited flat bands are accessible through the numerically efficient formulation of Eq.~(\ref{eq11}). Finally, we demonstrate two distinctive transmission properties of the electrically switchable flat bands: step-like increased Hall conductance and the strongly enhanced density of states at the specific sequence of electric field strengths.

Beyond the 2DEG platform, the drift-cancellation mechanism introduced here can find a natural extension in strain-engineered graphene, where the inhomogeneous strain generates a pseudo-magnetic-field dipole profile \cite{Chcnature}. This field is expected to produce valley-polarized quantized Hall currents with opposite signs in the two valleys. This provides an interesting direction for future investigation, while experimental realization would require suitable strain engineering and corresponding measurement schemes. 


By linking classical drift cancellation, exact quantum solvability, and distinct transport signatures within a single field-controlled framework, our results demonstrate an electrically switchable mechanism for flat-band engineering in low-dimensional systems and provide a versatile platform for exploring quantized Hall current controlled by tuning source-drain voltage.

\section*{Acknowledgements} 
We gratefully thank Carmine Ortix for the valuable discussions. This work was supported by the National Science and Technology Council (Grant numbers NSTC 114-2112-M-001-059, 112-2112-M-006-026-, 112-2112-M-004-007 and 112-2112-M-006-015-MY2), and Academia Sinica (ASTP-113-M02) This work was supported in part by the Higher Education Sprout Project, Ministry of Education to the Headquarters of University Advancement at the National Cheng Kung University.
\bibliographystyle{apsrev4-2}
\bibliography{refs}

@inbook{superconductivity1,
author = {Hott, Roland and Kleiner, Reinhold and Wolf, Thomas and Zwicknagl, Gertrud},
publisher = {Wiley, New York},
isbn = {9783527600434},
title = {Review on Superconducting Materials},
booktitle = {Encyclopedia of Applied Physics},
chapter = {},
pages = {1--55},
doi = {10.1002/3527600434.eap790},
url = {https://onlinelibrary.wiley.com/doi/abs/10.1002/3527600434.eap790},
year = {2016}
}

@article{superconductivity2,
  author  = {Zhou, Xingjiang and Lee, Wei-Sheng and Imada, Masatoshi and
             Trivedi, Nandini and Phillips, Philip and Kee, Hae-Young and
             T{\"o}rm{\"a}, P{\"a}ivi and Eremets, Mikhail},
  title   = {High-temperature superconductivity},
  journal = {Nat. Rev. Phys.},
  volume  = {3},
  number  = {7},
  pages   = {462--465},
  year    = {2021},
  doi     = {10.1038/s42254-021-00324-3},
}

@article{superconductivity3,
title = {A review of recent advancement in superconductors},
journal = {Mater. Today Proc.},
volume = {37},
pages = {3612-3614},
year = {2021},
issn = {2214-7853},
doi = {10.1016/j.matpr.2020.09.771},
url = {https://www.sciencedirect.com/science/article/pii/S2214785320375234},
author = {Harvinderjeet Kaur and Harmanpreet Kaur and Anjana Sharma},
}

@misc{Hall1,
 title={Lectures on the Quantum Hall Effect}, 
 author={David Tong},
 eprint={1606.06687},
 archivePrefix={arXiv}, 
 year={2016},
}

@article{Hall2,
  title = {Resonant nonlinear Hall effect in two-dimensional electron systems},
  author = {Huang, Botsz and Moghaddam, Ali G. and Facio, Jorge I. and Chang, Ching-Hao},
  journal = {Phys. Rev. B},
  volume = {104},
  issue = {16},
  pages = {165303},
  numpages = {9},
  year = {2021},
  month = {Oct},
  publisher = {American Physical Society},
  doi = {10.1103/PhysRevB.104.165303},
  url = {https://link.aps.org/doi/10.1103/PhysRevB.104.165303}
}

@article{Hall3,
  title = {On-demand higher-harmonic generation through nonlinear Hall effects in curved nanomembranes},
  author = {Huang, Botsz and Huang, You-Ting and Yang, Jan-Chi and Chen, Tse-Ming and Moghaddam, Ali G. and Chang, Ching-Hao},
  journal = {Phys. Rev. B},
  volume = {109},
  issue = {13},
  pages = {134419},
  numpages = {7},
  year = {2024},
  month = {Apr},
  publisher = {American Physical Society},
  doi = {10.1103/PhysRevB.109.134419},
  url = {https://link.aps.org/doi/10.1103/PhysRevB.109.134419}
}

@article{BoltzHall,
title = {High harmonic Hall currents driven by curved conducting nanoarchitecture},
journal = {Mater. Today Phys.},
volume = {60},
pages = {101965},
year = {2026},
issn = {2542-5293},
doi = {10.1016/j.mtphys.2025.101965},
url = {https://www.sciencedirect.com/science/article/pii/S2542529325003219},
author = {Botsz Huang and Wei-Xiang Yin and Xiao Zhang and Ching-Hao Chang},
}

@article{information1,
   title={Quantum Information},
   volume={139},
   ISSN={0587-4246},
   url={http://dx.doi.org/10.12693/APhysPolA.139.197},
   DOI={10.12693/aphyspola.139.197},
   number={3},
   journal={Acta Phys. Pol. A},
   publisher={Institute of Physics, Polish Academy of Sciences},
   author={Horodecki, R.},
   year={2021},
   month=mar, pages={197–2018} }

@article{information2,
  author  = {Zapatero, Víctor and van Leent, Tim and Arnon-Friedman, Rotem and Liu, Wen-Zhao and Zhang, Qiang and Weinfurter, Harald and Curty, Marcos},
  title   = {Advances in device-independent quantum key distribution},
  journal = {npj Quantum Inf.},
  volume  = {9},
  pages   = {10},
  year    = {2023},
  doi     = {10.1038/s41534-023-00684-x}
}

@article{flatband1,
  author  = {Checkelsky, Joseph G. and Bernevig, B. Andrei and Coleman, Piers and Si, Qimiao and Paschen, Silke},
  title   = {Flat bands, strange metals and the Kondo effect},
  journal = {Nat. Rev. Mater.},
  volume  = {9},
  pages   = {509--526},
  year    = {2024},
  doi     = {10.1038/s41578-023-00644-z}
}

@article{flatband2,
author = {Derzhko, Oleg and Richter, Johannes and Maksymenko, Mykola},
title = {Strongly correlated flat-band systems: The route from Heisenberg spins to Hubbard electrons},
journal = {Int. J. Mod. Phys. B},
volume = {29},
number = {12},
pages = {1530007},
year = {2015},
doi = {10.1142/S0217979215300078},

URL = {https://doi.org/10.1142/S0217979215300078
}
}

@article{flatband3,
   title={Artificial flat band systems: from lattice models to experiments},
   volume={3},
   ISSN={2374-6149},
   url={http://dx.doi.org/10.1080/23746149.2018.1473052},
   DOI={10.1080/23746149.2018.1473052},
   number={1},
   journal={Adv. Phys. X},
   publisher={Informa UK Limited},
   author={Leykam, Daniel and Andreanov, Alexei and Flach, Sergej},
   year={2018},
   month=jan, pages={1473052} }

@book{Kittel,
  author    = {Kittel, Charles},
  title     = {Introduction to Solid State Physics},
  publisher = {Wiley},
  address   = {New York},
  year      = {2005},
  edition   = {8th}
}

@article{flatband4,
  title = {Exact Landau Level Description of Geometry and Interaction in a Flatband},
  author = {Wang, Jie and Cano, Jennifer and Millis, Andrew J. and Liu, Zhao and Yang, Bo},
  journal = {Phys. Rev. Lett.},
  volume = {127},
  issue = {24},
  pages = {246403},
  numpages = {6},
  year = {2021},
  month = {Dec},
  publisher = {American Physical Society},
  doi = {10.1103/PhysRevLett.127.246403},
  url = {https://link.aps.org/doi/10.1103/PhysRevLett.127.246403}
}

@article{flatband5,
  author  = {Chen, Lei and Xie, Fang and Sur, Shouvik and Hu, Haoyu and Paschen, Silke and Cano, Jennifer and Si, Qimiao},
  title   = {Emergent flat band and topological Kondo semimetal driven by orbital-selective correlations},
  journal = {Nat. Commun.},
  volume  = {15},
  pages   = {5242},
  year    = {2024},
  doi     = {10.1038/s41467-024-49306-w}
}

@article{flatband6,
  title = {Flat bands in slightly twisted bilayer graphene: Tight-binding calculations},
  author = {Su\'arez Morell, E. and Correa, J. D. and Vargas, P. and Pacheco, M. and Barticevic, Z.},
  journal = {Phys. Rev. B},
  volume = {82},
  issue = {12},
  pages = {121407},
  numpages = {4},
  year = {2010},
  month = {Sep},
  publisher = {American Physical Society},
  doi = {10.1103/PhysRevB.82.121407},
  url = {https://link.aps.org/doi/10.1103/PhysRevB.82.121407}
}

@article{Magicfield,
  title = {Emergent flat band lattices in spatially periodic magnetic fields},
  author = {Tahir, M. and Pinaud, Olivier and Chen, Hua},
  journal = {Phys. Rev. B},
  volume = {102},
  issue = {3},
  pages = {035425},
  numpages = {9},
  year = {2020},
  month = {Jul},
  publisher = {American Physical Society},
  doi = {10.1103/PhysRevB.102.035425},
  url = {https://link.aps.org/doi/10.1103/PhysRevB.102.035425}
}

@article{Danieli2021,
author = {C. Danieli and S. Flach},
title = {Progress on artificial flat band systems: classifying, perturbing, applying},
journal = {Advances in Physics},
volume = {0},
number = {0},
pages = {1--10},
year = {2026},
publisher = {Taylor \& Francis},
doi = {10.1080/00018732.2026.2658992},
url ={https://doi.org/10.1080/00018732.2026.2658992}
}

@article{Danieli2024,
author = {Danieli, Carlo and Andreanov, Alexei and Leykam, Daniel and Flach, Sergej},
title = {Flat band fine-tuning and its photonic applications},
journal = {Nanophotonics},
volume = {13},
number = {21},
pages = {3925-3944},
url = {https://onlinelibrary.wiley.com/doi/abs/10.1515/nanoph-2024-0135},
year = {2024}
}

@article{Poblete2021,
author = {Rodrigo A. Vicencio Poblete},
title = {Photonic flat band dynamics},
journal = {Advances in Physics: X},
volume = {6},
number = {1},
pages = {1878057},
year = {2021},
publisher = {Taylor \& Francis},
doi = {10.1080/23746149.2021.1878057},
URL = {https://doi.org/10.1080/23746149.2021.1878057},
}

@article{flatband7,
author = {Jeronimo G. C. Martinez  and Christie S. Chiu  and Basil M. Smitham  and Andrew A. Houck },
title = {Flat-band localization and interaction-induced delocalization of photons},
journal = {Sci. Adv.},
volume = {9},
number = {50},
pages = {eadj7195},
year = {2023},
doi = {10.1126/sciadv.adj7195},
URL = {https://www.science.org/doi/abs/10.1126/sciadv.adj7195}
}

@book{griffiths_introduction_2018,
	address = {Cambridge},
	edition = {3},
	title = {Introduction to Quantum Mechanics},
	isbn = {978-1-107-18963-8},
	publisher = {Cambridge University Press},
	author = {Griffiths, David J. and Schroeter, Darrell F.},
	year = {2018},
}

@article{PhysRevLett1992,
  title = {Effect of a nonuniform magnetic field on a two-dimensional electron gas in the ballistic regime},
  author = {M\"uller, J. E.},
  journal = {Phys. Rev. Lett.},
  volume = {68},
  issue = {3},
  pages = {385--388},
  numpages = {0},
  year = {1992},
  month = {Jan},
  publisher = {American Physical Society},
  doi = {10.1103/PhysRevLett.68.385},
  url = {https://link.aps.org/doi/10.1103/PhysRevLett.68.385}
}

@article{ChingHao,
  author = {C. H. Chang and J. van den Brink and C. Ortix},
  title = {Strongly anisotropic ballistic magnetoresistance in compact three-dimensional semiconducting nanoarchitectures},
  journal = {Phys. Rev. Lett.},
  volume = {113},
  pages = {227205},
  year = {2014},
  doi = {10.1103/PhysRevLett.113.227205}
}

@article{ChangHao2,
  author  = {C. H. Chang and C. Ortix},
  title   = {Theoretical Prediction of a Giant Anisotropic Magnetoresistance in Carbon Nanoscrolls},
  journal = {Nano Lett.},
  year    = {2017},
  date    = {2017-05-10},
  volume  = {17},
  number  = {5},
  pages   = {3076--3080},
  doi     = {10.1021/acs.nanolett.7b00426},
}

@book{plasma,
author = {Chen, Francis},
year = {2016},
month = {01},
title = {Introduction to Plasma Physics and Controlled Fusion},
isbn = {978-3-319-22308-7},
publisher = {Springer, Cham},
doi = {10.1007/978-3-319-22309-4}
}

@article{anharmonic,
  title = {Anharmonic Oscillator},
  author = {Bender, Carl M. and Wu, Tai Tsun},
  journal = {Phys. Rev.},
  volume = {184},
  issue = {5},
  pages = {1231--1260},
  numpages = {0},
  year = {1969},
  month = {Aug},
  publisher = {American Physical Society},
  doi = {10.1103/PhysRev.184.1231},
  url = {https://link.aps.org/doi/10.1103/PhysRev.184.1231}
}

@article{2DEG1,
  author = {W. T. Sommer},
  title = {Liquid helium as a barrier to electrons},
  journal = {Phys. Rev. Lett.},
  volume = {12},
  pages = {271},
  year = {1964},
  doi = {10.1103/PhysRevLett.12.271}
}

@article{2DEG2,
  author = {K. v. Klitzing and G. Dorda and M. Pepper},
  title = {New method for high-accuracy determination of the fine-structure constant based on quantized Hall resistance},
  journal = {Phys. Rev. Lett.},
  volume = {45},
  pages = {494},
  year = {1980},
  doi = {10.1103/PhysRevLett.45.494}
}

@incollection{ballistic,
title = {Quantum Transport in Semiconductor Nanostructures},
editor = {Henry Ehrenreich and David Turnbull},
series = {Solid State Physics},
publisher = {Academic Press},
volume = {44},
pages = {1-228},
year = {1991},
booktitle = {Semiconductor Heterostructures and Nanostructures},
issn = {0081-1947},
doi = {https://doi.org/10.1016/S0081-1947(08)60091-0},
url = {https://www.sciencedirect.com/science/article/pii/S0081194708600910},
author = {C.W.J. Beenakker and H. {van Houten}}
}

@misc{note1,
    key = {},
    note = {when the applied electric field magnitude is 
    $eV_e=1\hbar\omega_c$, 2 lowest bands become nondegenerate, while the rests recover degenerate (as in Fig. 3(c)); for $eV_e=2\hbar\omega_c$, 4 lowest bands become nondegenerate, and so on}
}

@book{math_for_physicist,
  author    = {Riley, K. F. and Hobson, M. P. and Bence, S. J.},
  title     = {Mathematical Methods for Physics and Engineering},
  publisher = {Cambridge University Press},
  address   = {Cambridge},
  year      = {2006},
  edition   = {3}
}

@article{Landauer,
    author = {Landauer, R.},
    title = {Spatial variation of currents and fields due to localized scatterers in metallic conduction (and comment)},
    journal = {J. Math. Phys.},
    volume = {37},
    number = {10},
    pages = {5259-5268},
    year = {1996},
    month = {10},
    issn = {0022-2488},
    doi = {10.1063/1.531590},
    url = {https://doi.org/10.1063/1.531590},
}

@article{Dweiss1997,
    author = {Ye, P. D. and Weiss, D. and Gerhardts, R. R. and Nickel, H.},
    title = {Magnetoresistance oscillations induced by periodically arranged micromagnets (invited)},
    journal = {Journal of Applied Physics},
    volume = {81},
    number = {8},
    pages = {5444-5448},
    year = {1997},
    month = {04},
    issn = {0021-8979},
    doi = {10.1063/1.364565},
    url = {https://doi.org/10.1063/1.364565},
}

@article{Roukes1987,
  title = {Quenching of the Hall Effect in a One-Dimensional Wire},
  author = {Roukes, M. L. and Scherer, A. and Allen, S. J. and Craighead, H. G. and Ruthen, R. M. and Beebe, E. D. and Harbison, J. P.},
  journal = {Phys. Rev. Lett.},
  volume = {59},
  issue = {26},
  pages = {3011--3014},
  numpages = {0},
  year = {1987},
  month = {Dec},
  publisher = {American Physical Society},
  doi = {10.1103/PhysRevLett.59.3011},
  url = {https://link.aps.org/doi/10.1103/PhysRevLett.59.3011}
}

@article{Dweiss1995,
  title = {Electrons in a Periodic Magnetic Field Induced by a Regular Array of Micromagnets},
  author = {Ye, P. D. and Weiss, D. and Gerhardts, R. R. and Seeger, M. and von Klitzing, K. and Eberl, K. and Nickel, H.},
  journal = {Phys. Rev. Lett.},
  volume = {74},
  issue = {15},
  pages = {3013--3016},
  numpages = {0},
  year = {1995},
  month = {Apr},
  publisher = {American Physical Society},
  doi = {10.1103/PhysRevLett.74.3013},
  url = {https://link.aps.org/doi/10.1103/PhysRevLett.74.3013}
}

@article{expB,
  title = {Transport in a two-dimensional electron-gas narrow channel with a magnetic-field gradient},
  author = {Hara, Masahiro and Endo, Akira and Katsumoto, Shingo and Iye, Yasuhiro},
  journal = {Phys. Rev. B},
  volume = {69},
  issue = {15},
  pages = {153304},
  numpages = {4},
  year = {2004},
  month = {Apr},
  publisher = {American Physical Society},
  doi = {10.1103/PhysRevB.69.153304},
  url = {https://link.aps.org/doi/10.1103/PhysRevB.69.153304}
}

@article{expE,
author = {J., Cyril Robinson Azariah and Rajesh, Swaminathan},
year = {2017},
month = {10},
pages = {613-623},
title = {Current trends in changing the channel in MOSFETs by III-V semiconducting nanostructures},
volume = {6},
journal = {Nanotechnol. Rev.},
doi = {10.1515/ntrev-2017-0155}
}

@article{Chcnature,
  author    = {Sheng-Chin Ho and Ching-Hao Chang and Yu-Chiang Hsieh and Shun-Tsung Lo and Botsz Huang and Thi-Hai-Yen Vu and Carmine Ortix and Tse-Ming Chen},
  title     = {Hall effects in artificially corrugated bilayer graphene without breaking time-reversal symmetry},
  journal   = {Nature Electronics},
  year      = {2021},
  volume    = {4},
  number    = {2},
  pages     = {116--125},
  doi       = {10.1038/s41928-021-00537-5}
}

@article{PRB_205309,
  title = {Giant asymmetry of the longitudinal magnetoresistance in high-mobility two-dimensional electron gas on a cylindrical surface},
  author = {Vorob'ev, A. B. and Friedland, K.-J. and Kostial, H. and Hey, R. and Jahn, U. and Wiebicke, E. and Yukecheva, Ju. S. and Prinz, V. Ya.},
  journal = {Phys. Rev. B},
  volume = {75},
  issue = {20},
  pages = {205309},
  numpages = {6},
  year = {2007},
  month = {May},
  publisher = {American Physical Society},
  doi = {10.1103/PhysRevB.75.205309},
  url = {https://link.aps.org/doi/10.1103/PhysRevB.75.205309}
}

\end{document}